\UseRawInputEncoding % Add this line to specify UTF-8 encoding
\documentclass[sn-mathphys,Numbered]{sn-jnl}% Math and Physical Sciences Reference Style
\usepackage{graphicx}%
\usepackage{multirow}%
\usepackage{amsmath,amssymb,amsfonts}%
\usepackage{amsthm}%
\usepackage{mathrsfs}%
\usepackage[title]{appendix}%
\usepackage{xcolor}%
\usepackage{textcomp}%
\usepackage{manyfoot}%
\usepackage{booktabs}%
\usepackage{algorithm}%
\usepackage{algorithmicx}%
\usepackage{algpseudocode}%
\usepackage{listings}%
\usepackage{graphicx}%
\usepackage{subcaption}%
\theoremstyle{thmstyleone}%
\theoremstyle{thmstyletwo}%

\theoremstyle{thmstylethree}%

\begin{document}

\title[Article Title]{The Impact of Surface Passivation on Kapitza Resistance at the Interface between a Semiconductor and Liquid Nitrogen}


\author*{\fnm{Babak} \sur{Mohammadian}*}\email{Babak.mohammadian@postgrad.manchester.ac.uk}
\author{\fnm{Mark A.} \sur{McCulloch}}
\author{\fnm{Thomas} \sur{Sweetnam}}
\author{\fnm{Valerio} \sur{Gilles}}
\author{\fnm{Lucio} \sur{Piccirillo}}

\affil{\orgdiv{Jodrell Bank Centre for Astrophysics}, \orgname{University of Manchester}, \orgaddress{\city{Manchester}, \country{UK}}}


\abstract{Cooling electronic devices to cryogenic temperatures ($<$ 77 K) is crucial in various scientific and engineering domains. Efficient cooling involves the removal of heat generated from these devices through thermal contact with either a liquid cryogen or a dry cryostat cold stage. However, as these devices cool, thermal boundary resistance, also known as Kapitza resistance, hinders the heat flow across thermal interfaces, resulting in elevated device temperatures. In transistors, the presence of passivation layers like Silicon Nitride (SiN) introduces additional interfaces that further impede heat dissipation. This paper investigates the impact of passivation layer thickness on Kapitza resistance at the interface between a solid device and liquid nitrogen. The Kapitza resistance is measured using a capacitance thermometer that has been passivated with SiN layers ranging from 0 to 240 nm. We observe that Kapitza resistance increases with increasing passivation thickness.}



\maketitle
 \vskip-2.5em
\section{Introduction}\label{sec1}
To ensure the linearity and reliability of electronic devices operating at high power and frequency, effective cooling is required \cite{b1,b2,b3}. In transistors, insufficient heat dissipation in the active channel leads to significant temperature elevations, commonly referred to as self-heating, which adversely impacts the device's performance \cite{b4,b5}. 
Intensive efforts have been undertaken to effectively understand and mitigate the challenge of self-heating, particularly at cryogenic temperature \cite{b6,b7,b8,b9}. For example, the simulations reported in \cite{b10} show that heat dissipation from the active channel of a transistor is limited by the lack of states in phonon black-body radiation. Consequently, even when devices are cooled below 1 K, the temperature within the active region remains 10 to 20 K. For transistors, this phenomenon causes the noise figure to plateau below 20 K \cite{b101}.\\ 
Further, to avoid damage and contamination, transistors are commonly coated with a nitride or oxide passivation layer. It is reported in \cite{b19} that at room temperature, the self-heating in the active channel of the power transistor is affected by the passivation layer's thickness and the thermal conductivity of materials utilized in the device. However, the effect of surface passivation on thermal dispersion has not been measured at cryogenic temperatures. In this paper, we preliminary present the results showing the impact of passivation layer thickness on the Kapitza resistance between a gold-plated quartz substrate  and liquid nitrogen. In section \ref{sec2}, we will discuss Kapitza resistance and how it can impede phonon transmission at material interfaces. In sections \ref{sec3} and \ref{sec:calibration}, capacitor fabrication and thermometer calibration are presented, and related plots are discussed. Finally, our experimental setup and how we use a capacitance thermometer to measure the Kapitza resistance are presented in section \ref{sec5}. 
\section{Kapitza Resistance}\label{sec2}
When heat flows from a solid device to a liquid cryogen, there will be a temperature difference in the interfacial area. This temperature difference is due to the acoustic impedance mismatch between the materials, which impedes the flow of the phonons across the interface, resulting in a measurable resistance known as Kapitza resistance \cite{b20,b24,b201}. The Kapitza resistance, $R_K$, for constant thermal flux at the interface is given by:

\begin{equation}
R_{K} = A \cdot \frac{{\Delta{T}}}{{\dot{Q}}}  
 (\mathrm{m}^2\mathrm{K/W})
\label{eq:1}
\end{equation}
where $\Delta{T}$ is the temperature difference across the liquid-solid interface, $A$ is the area of the interface, and $\dot{Q}$ is the total heat flow.\\
\noindent In the case of transistors, when transistors are cooled with liquid cryogens, we hypothesize that the Kapitza resistance hinders the phonon heat dissipation from the active region, resulting in it being hotter than its surroundings \cite{b10}.

\section{The Fabrication}\label{sec3}
\begin{figure}[h!]
\centering
\includegraphics[width=10cm]{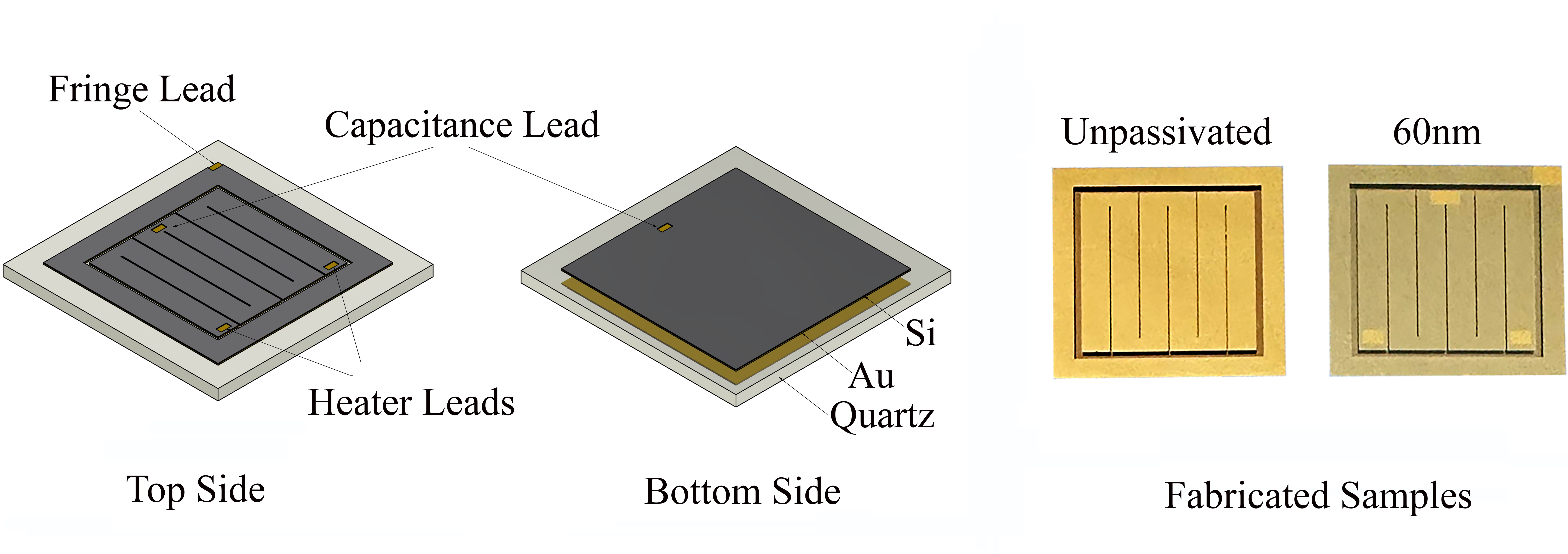}
\caption{The capacitor model, consisting of a quartz dielectric,  gold plates, an embedded heater, and different thicknesses of SiN (top and bottom sides). Fabricated samples: Unpassivated and 60nm of SiN. The 3 small rectangles in the 60 nm samples are unpassivated to allow the capacitance measurement and bias wires to be attached.}
\label{fig:wafer_model}
\end{figure}

To investigate the effect of the surface passivation layer on the Kapitza resistance at the interface of the sample and liquid nitrogen, we have fabricated parallel plate quartz capacitors that act as thermometers \cite{b25}. Unlike certain materials that experience significant fluctuations in dielectric constants ($\varepsilon$) with temperature, quartz provides a stable dielectric constant as well as high hardness and minimal thermal contraction stability under temperature variation \cite{bquartz}. Also, since its dielectric constant is temperature-dependent in the region of interest (60-100 K), it is chosen as the dielectric of the capacitor in this experiment. To fabricate the capacitors, the quartz wafer is diced into $20\times20$
 \text {mm}$^2$ chips with the substrate thickness tolerance and cutting deviation of  $\pm$20 $\mu$m and  $\pm$0.1 mm, respectively. The samples are then patterned with a laser writer and coated with (50 $\pm$ 2) nm layers of Au on both sides, using an Electron Beam Evaporator. Further, to vary the heat flux across the interface of Au-SiN, one side of the capacitor acts as a heater, using a meandering path that is patterned with Photo-lithography and a Lift-off process \cite{b25}. Also, a thin border electrode (guard electrode) is used to prevent the chip from stray and fringe capacitances. In the second phase of fabrication, different passivation layers of $\text{SiN}$, ranging 0 $\sim$ 240 nm, are deposited over both sides of the capacitor by Plasma-Enhanced Chemical Vapor Deposition (PECVD). Fig. \ref{fig:wafer_model} illustrates the fabricated samples.
 \vskip-1.6em
 \vskip-1.2em
 \noindent To use the fabricated capacitor as a reliable thermometer, it needs to be calibrated. Therefore, the variation of the dielectric constant of quartz should be measured with respect to temperature to provide a calibration plot. 

 \section{Calibration}\label{sec:calibration}
\begin{figure}[h!]
\centering
\includegraphics[width=6.5 cm]{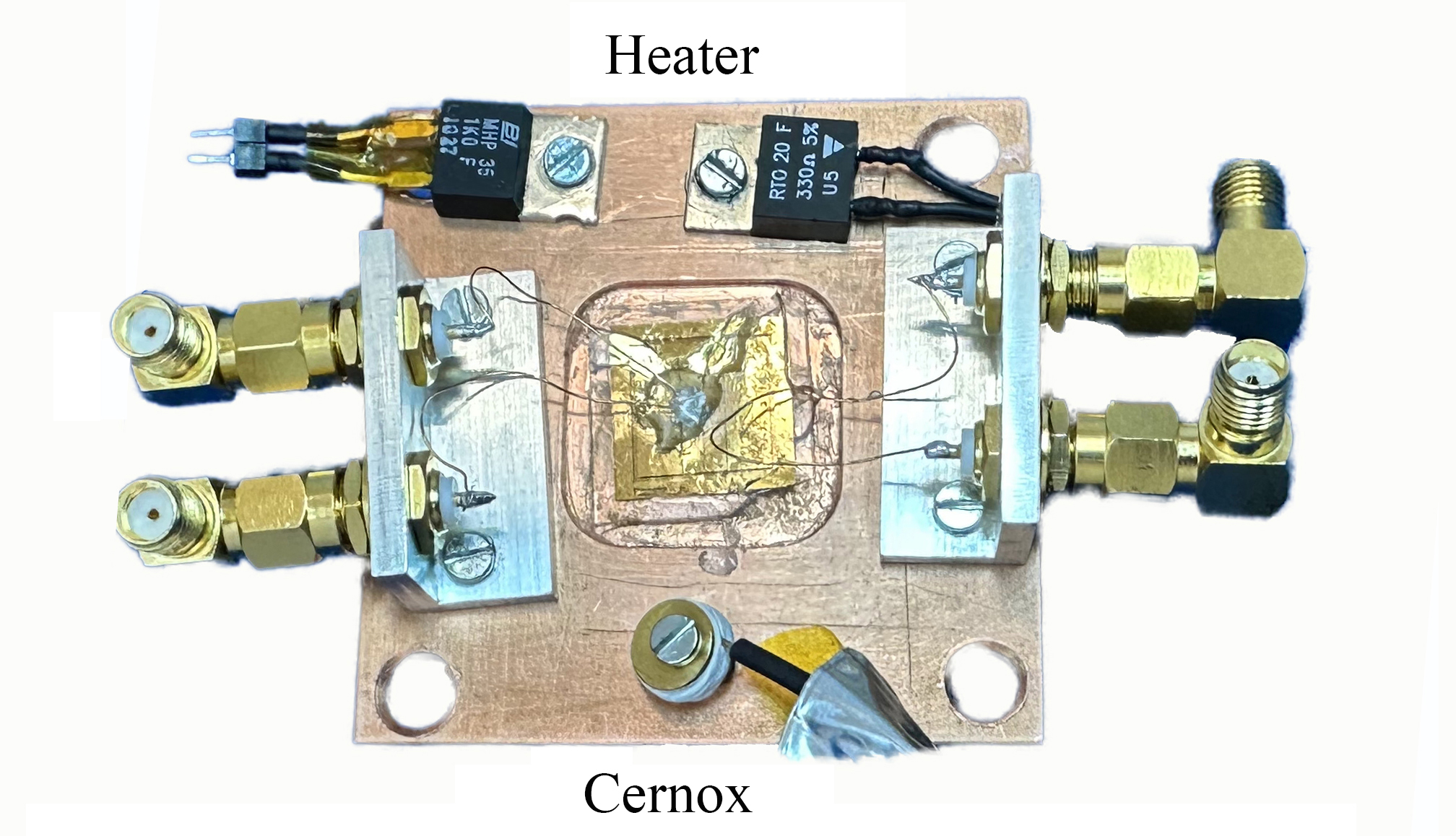}
\caption{The copper sample holder, with the central sample epoxied in place and connected to the outer SMA connectors with copper wires. The heaters allow the temperature of the sample to be varied, with the calibrated thermometer (Cernox) providing an accurate temperature value.}
 \label{fig:sample_holder}
\end{figure}

A cryostat equipped with a two-stage mechanical cooler (dry cooling) is used to calibrate the dielectric constant of the samples against temperatures from 60 to 100 K. A copper sample holder is used to provide a thermal contact for the sample in the cooling process, Fig. \ref{fig:sample_holder}.\\It contains two heaters, a calibrated temperature sensor (Lakeshore Cernox: CX-1030-CU-HT-0.3L). To measure the capacitance, two pairs of 0.1 mm-thick shielded copper wires are indium soldered at the allocated bond pads on the top and bottom of the samples. The copper wires are then soldered to the SMA connectors, which are connected to a Keysight LCR meter Model: E4980AL using stainless steel coaxial cables.  It is then attached to the 4 K stage of a cryostat to conduct the calibration process.\\

\begin{figure}[ht!]
    \centering
    \begin{subfigure}[t]{0.48\textwidth}
        \centering
        \includegraphics[width=\linewidth]{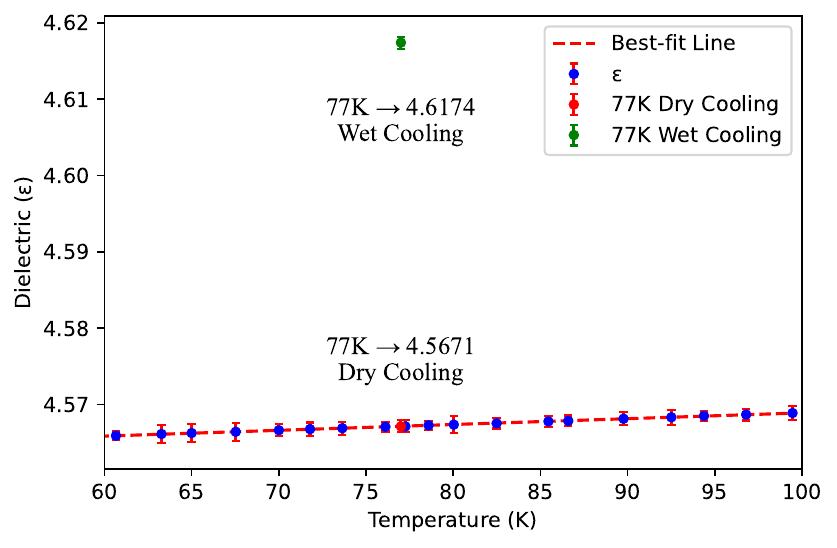}
        \caption{}
        \label{fig:dry_cooling}
    \end{subfigure}\hfill
    \begin{subfigure}[t]{0.5\textwidth}
        \centering
        \includegraphics[width=\linewidth]{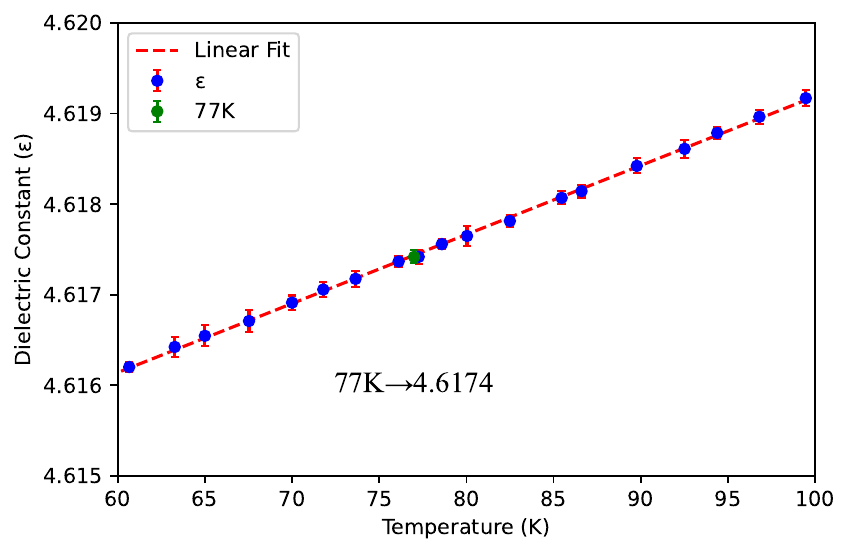}
        \caption{}
        \label{fig:shifted}
    \end{subfigure}
    \caption{ Calibration plot: (a)  The calibration plot in dry cooling (cryostat). The red marker shows the measured dielectric constant at 77 K (the reference point in dry cooling), while the green marker illustrates the measured dielectric constant in liquid nitrogen (77K). (b) Adjusted calibration plot from 60K to 100K due to the change in measuring environment (liquid nitrogen reference point).}
    \label{fig:calibration_plots}
\end{figure}

\newpage 
\noindent  The calibration plots for cooling are shown in Fig. \ref{fig:calibration_plots}. Since the experiment is based on the immersion of samples in liquid nitrogen, the acquired calibration data requires a correction to compensate for the slightly different electromagnetic environment seen by the capacitor when immersed in the liquid nitrogen dewar.

\vskip -0.5em
\noindent Fig. \ref {fig:dry_cooling} shows that the dielectric constant in liquid nitrogen (wet cooling) is higher than that measured in the cryostat (dry cooling). Therefore, we apply a linear shift to the calibration to correct for the change in the environment, which can be seen in Fig. \ref {fig:shifted}. In order to validate this phenomenon, we conducted an experiment involving the immersion of five separate samples (ranging from 0 to 240 nm) in liquid nitrogen and measured the capacitance using LCR and averaged. The effect was repeatable with the measured dielectric constant at 4.61741 $\pm$ 7.78E-05 and assumed to be linear across the temperature region of interest.

\section{Kapitza Resistance Measurement}\label{sec5}
To derive the Kapitza resistance using  Eq.\ref{eq:1}, it is required to measure the surface temperature while applying constant power for different thicknesses of passivation. 

 \begin{figure}[ht]
    \centering
    \begin{subfigure}[b]{0.35\textwidth}
        \centering
        \includegraphics[width=0.65\linewidth]{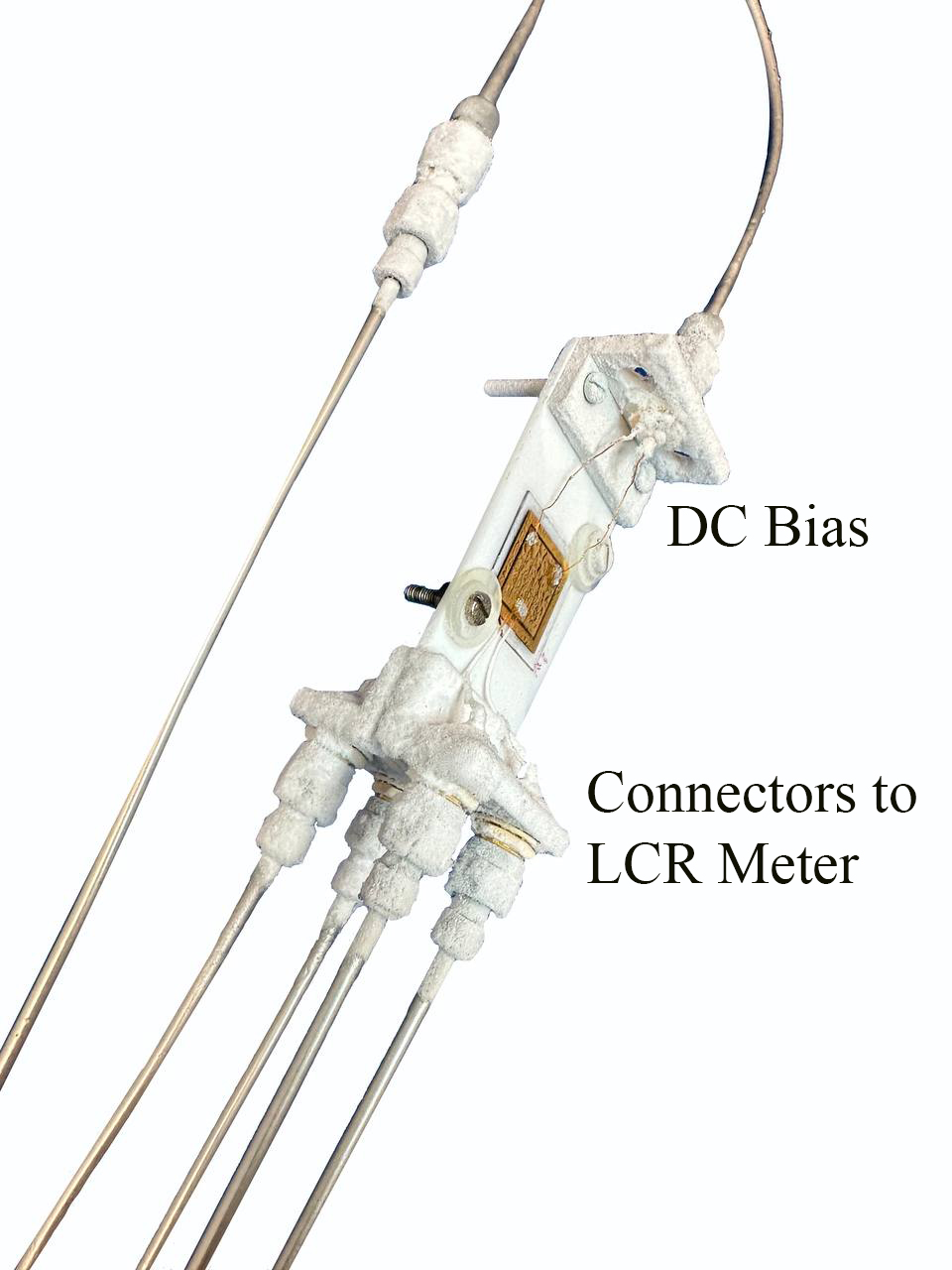}
        \caption{}
        \label{fig:bucket}
    \end{subfigure}
    \begin{subfigure}[b]{0.35\textwidth}
        \centering
        \includegraphics[width=\linewidth]{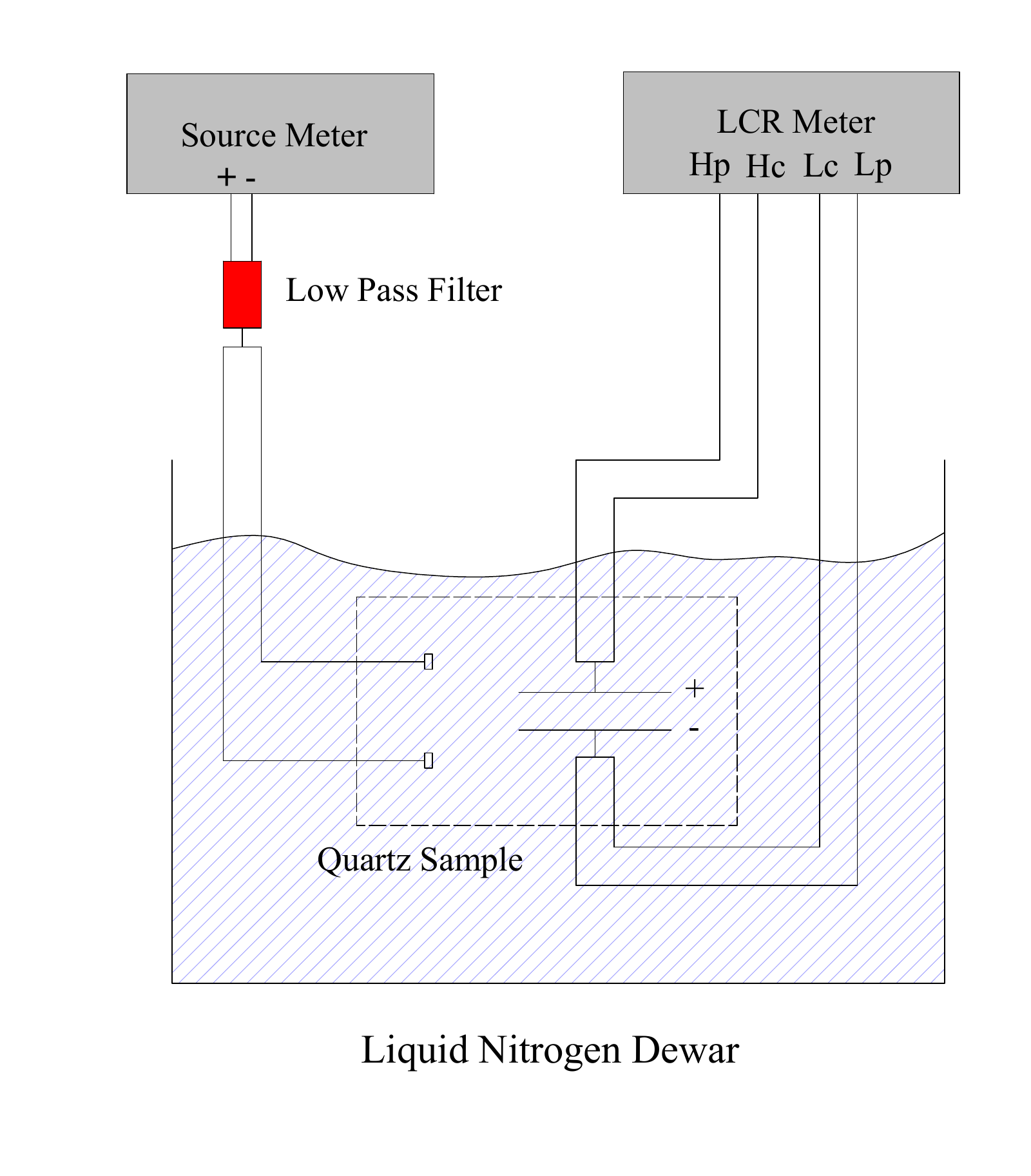} % Adjust the width as desired
        \vskip -0.5cm
        \caption{}
        \label{fig:sma}
    \end{subfigure}
    \vskip -0.5em
    \caption{(a) PTFE-based sample holder. (b) Measurement setup, comprising the measurement schematic of 4-wire measurement of capacitance and DC biasing of the heater  }
    \label{fig:setup}
    \hspace{2em}
    
\end{figure}
\begin{figure}[ht]
\centering
\includegraphics[width=9cm]{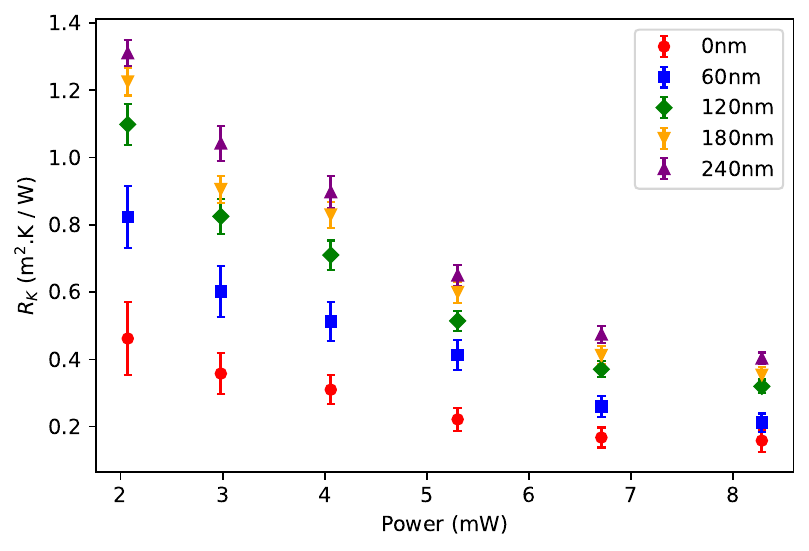}

\caption{Kapitza resistance values for different passivation layers and applied power. Considering a constant power for each sample, by increasing the thickness of the passivation layer, the Kapitza resistance increases.} 
\label{fig: Kapitza Measurement}
\end{figure}

\vskip -1.5em
\noindent To minimize noise and the impact of the test fixture, a 4-wire capacitance measurement, as explained in Section \ref{sec:calibration}, is performed.  Also, to provide the DC bias, two bond pads are allocated at each end of the meander and connected to a Keithley source meter, model: 2401, via coaxial cables. In this way, the applied DC current makes the meander a controllable heater ($\sim$ 14 $\Omega$). The wired sample is then placed on a Polytetrafluoroethylene (PTFE) sample holder  (thickness: 1.5 mm), using plastic washers Fig. \ref{fig:bucket}. 
 A low pass filter with a cutoff frequency of 100 kHz is also used to reduce unwanted noise interference (mostly from the source meter at 80 kHz). The set-up can be seen in \ref{fig:sma}.\\ 
To measure the Kapitza resistance, each sample undergoes immersion in the liquid nitrogen bath for 20 minutes to ensure it is properly thermalized. The volume of the liquid nitrogen is chosen in a way that applying heat to the sample does not warm the reservoir significantly. Also, the generated heat by the LCR meter into the sample is in the order of nW, which is negligible.\\  For each sample, 200 readings of the capacitance at a frequency of 200 kHz are acquired and averaged. This is repeated at six different power levels, ranging from 2.04 mW to 8.28 mW, corresponding to a current sweep from 10 to 20 mA in steps of 2 mA. The dielectric constant is then calculated, and the corresponding temperature value is found from the fit in  Fig. \ref{fig:shifted}. The Kapitza resistance is then calculated using Eq. \ref{eq:1} and shown in Fig \ref{fig: Kapitza Measurement}.\\

 \vskip -1em
\noindent The data in Fig. \ref{fig: Kapitza Measurement} shows that for a constant power, as the passivation thickness increases, there is a corresponding rise in the Kapitza resistance. Therefore, this supports our assumption that the passivation layer has a direct impact on the self-heating of the active regions in cryogenic electronics.
%===================

\section{Conclusion}
Self-heating behavior of the active channel of transistors puts constraints on the performance of these devices. Since the surface of transistors is commonly coated with a passivation layer, thermal dissipation can be restricted by this layer. In this paper, we investigated an experimental setup to evaluate the effect of surface passivation on Kapitza resistance at the interface of a solid and liquid nitrogen. We found that the Kapitza resistance increased for thicker passivation layers, increasing the self-heating effect. The results suggest that since Kapitza resistance increases with the variation of the passivation layer thickness, reducing the passivation layer thickness to a desirable thickness may improve the self-heating in transistors, and further work is underway to investigate this. 
\newenvironment{acknowledgments}
  {\section*{Acknowledgments}}
  {}

\begin{acknowledgments}
This project received funding from the European Union’s Horizon 2020 research and innovation program under the Marie Skłodowska-Curie grant agreement No. 811312 for the project “Astro-Chemical Origins” (ACO) and UKRI ST/X006344/1.\\
We wish to acknowledge the support of the National Graphene Institute team, Dr. Lee Hague, Andrew Brook, Robert Howard, Matthew Whitelegg, and Dr. Kunal Lulla, offering suggestions in the fabrication process.

\end{acknowledgments}

\bibliography{sn-article}% common bib file

\end{document}